\documentclass[a4paper,12pt]{article}
\usepackage[utf8]{inputenc}
\usepackage[paper=a4paper,left=25mm,right=25mm,top=25mm,bottom=25mm]{geometry} 

\usepackage{setspace}
\usepackage{amsfonts}

\usepackage{amsmath,amssymb}

\usepackage{hyperref}
\hypersetup{colorlinks = TRUE,
linkcolor = magenta,
citecolor=blue}

\usepackage{enumitem}
\makeatletter
\def\namedlabel#1#2{\begingroup
	#2%
	\def\@currentlabel{#2}%
	\phantomsection\label{#1}\endgroup
}
\makeatother

\usepackage[dvipsnames]{xcolor}

\usepackage{enumitem}

\usepackage[normalem]{ulem}

\usepackage{graphicx}

\usepackage{tikz}
\usetikzlibrary{positioning, arrows.meta, tikzmark,fit,shapes.geometric}
\usepackage[
backend=biber,
style=apa,autocite=inline,
sorting=nyt
]{biblatex}

\usepackage{booktabs}

\usepackage{pdflscape}
\usepackage{afterpage}

\usepackage{authblk}

\title{Validation of cluster analysis results on validation data: A systematic framework}
\author[1]{Theresa Ullmann\thanks{Corresponding author, e-mail: \href{mailto:tullmann@ibe.med.uni-muenchen.de}{tullmann@ibe.med.uni-muenchen.de}, Institute for Medical Information Processing, Biometry, and Epidemiology, Ludwig-Maximilians-Universität München, Marchioninistr. 15, D-81377, Munich, Germany.}}
\author[2]{Christian Hennig}
\author[1]{Anne-Laure Boulesteix}
\affil[1]{Institute for Medical Information Processing, Biometry and Epidemiology, LMU Munich, Germany}
\affil[2]{Dipartimento di Scienze Statistiche ``Paolo Fortunati'', Universita di Bologna, Italy}
\date{March 1st, 2021}

\begin{document}

% !BIB TS-program = biber

\maketitle

\textbf{Please note:} This arXiv preprint is an older version of the manuscript. An updated and improved version has been published \textbf{open access} in WIREs Data Mining and Knowledge Discovery:

Ullmann, T., Hennig, C.,\& Boulesteix, A.-L. (2021). Validation of cluster analysis results on validation data: A systematic framework. Wiley Interdisciplinary Reviews: Data Mining and Knowledge Discovery, e1444. \textbf{\url{https://doi.org/10.1002/widm.1444}}

\newpage

\section*{Abstract}
Cluster analysis refers to a wide range of data analytic techniques for class discovery and is popular in many application fields. To judge the quality of a clustering result, different cluster validation procedures have been proposed in the literature. While there is extensive work on classical validation techniques, such as internal and external validation, less attention has been given to validating and replicating a clustering result using a validation dataset. Such a dataset may be part of the original dataset, which is separated before analysis begins, or it could be an independently collected dataset. We present a systematic structured framework for validating clustering results on validation data that includes most existing validation approaches. In particular, we review classical validation techniques such as internal and external validation, stability analysis, hypothesis testing, and visual validation, and show how they can be interpreted in terms of our framework.  We precisely define and formalise different types of validation of clustering results on a validation dataset and explain how each type can be implemented in practice. Furthermore, we give examples of how clustering studies from the applied literature that used a validation dataset can be classified into the framework.  

\section{Introduction}
\label{sec:intro}
Cluster analysis refers to data analytic techniques for class discovery. It is popular in a range of fields, for example,  medicine, biology, market research, social science, and data compression. However, when conducting cluster analysis, researchers are confronted with an overwhelming number of existing methods. They must preprocess the data, choose a clustering algorithm, and set parameters, such as the number of clusters (\cite{van_mechelen_benchmarking_2018}). It is often unclear a priori which choice should be made for the analysis, and even once a choice is made, it may remain unclear how good the quality of the resulting clustering is. 

These problems have prompted the development of so-called \textit{cluster validation} techniques, see \textcite{handl_computational_2005} and \textcite{hennig_clustering_2015} for overviews.  The literature mainly distinguishes between internal validation (where the clustering is evaluated based on internal properties, such as compactness and separateness of the clusters) and external validation (where the clustering is evaluated by comparing the clusters with respect to one or more variables not used for clustering, e.g., a survival time or a true class membership).  
Less attention has been given to the validation and replication of clustering results on a \textit{validation dataset}, for which we introduce a systematic framework. A validation dataset could be part of the original dataset, set apart before the start of the analysis, or it could be a separate dataset, obtained, for example, from a different study centre.

In the present paper, we precisely define and formalise the concepts of validation and replication on a validation dataset in different variants. We also discuss which aspects of the clustering process these terms refer to, which clustering properties can be validated, and how this can be implemented in practice. Moreover, we interpret various validation-related approaches proposed in the literature within our framework, which helps with the understanding of their potential uses, as well as advantages and limitations over alternative approaches. 

The idea of validating a clustering on another dataset is not new and has appeared in the methodological literature decades ago (\cite{mcintyre_nearest-centroid_1980,breckenridge_replicating_1989}). In applied literature involving cluster analysis, it is not uncommon for authors to validate their clustering results on new data, be it with the procedure of \textcite{mcintyre_nearest-centroid_1980} or another method. To the best of our knowledge, these approaches have never been systematically structured and evaluated. This contrasts with the abundant methodological literature devoted to validation in the context of \textit{supervised} classification (or more generally, supervised learning): the validation of such models on validation data is routinely performed and well-understood.
This contrast may be partly due to the fact that cluster analysis---as opposed to supervised classification---is often viewed as \lq\lq exploratory research''. The validation of  clustering results is rightly considered to be less straightforward than the validation of a prediction model because ``true labels'' are unknown (\cite{von_luxburg_clustering_2012}). Indeed, it is difficult to define exactly what is meant by validating a clustering on validation data. Answering this question is the key aspect of our framework.\\
There are multiple reasons why we consider a framework for validation and replication of clustering results on a validation dataset to be useful. Researchers usually want their results to be as generalisable as possible. This means that interesting properties of a clustering result would hold not only for a single specific dataset, but would also reappear when clustering validation data that is sampled from the same, or even different distributions. A validation framework may also enable researchers to evaluate clustering results reported by other research teams. The confirmation of results on validation data is a vital part of research in general, and it has received considerable attention in recent years due to the so-called ``replication crisis'' (\cite{hutson_artificial_2018}). 

Notably, the validation of clustering results on a validation dataset may also allow detection of ``overoptimism'' due to ``overfitting'' effects: When researchers try different clustering algorithms or parameters during the analysis, they can use classical internal and external validation methods to choose a single clustering out of these. However, the more  clustering methods tried, the more likely it is that one of them yields a satisfying result by chance---where the definition of the term ``satisfying'' depends on the context and on the researcher's subjective appraisal. Consequently, the reported results may be overoptimistic in the sense of reliability, similarly to the results of multiple tests if no adjustment is performed. While these mechanisms are well-understood in the context of multiple testing, this is less so in the context of clustering. Repeating the same cluster analysis on another dataset may be a sensible approach to ensure that seemingly satisfactory results are not (solely) the product of such overfitting effects. 

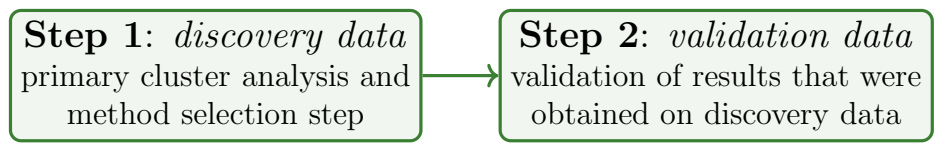
\begin{figure}[h]
	\label{fig:twostep}
	\centering
	\begin{tikzpicture}[
	textnode/.style={rectangle, rounded corners, draw=OliveGreen, fill=OliveGreen!5, very thick, minimum size=7mm, align=center},]
	\node[textnode]   (step1)   {{\large\textbf{Step 1}: \textit{discovery data}}\\
		primary cluster analysis and\\ method selection step};
	\node[textnode]   (step2)  [right=of step1] {{\large\textbf{Step 2}: \textit{validation data}}\\ validation of results that were\\ obtained on discovery data };
	\draw[->,draw=OliveGreen, very thick] (step1.east) -- (step2.west);
	\end{tikzpicture}
	\caption{Two-step procedure for validating clustering results.}
\end{figure}

Our framework is based on the following two-step cluster analysis procedure (see also Fig.~\ref{fig:twostep}): 
\begin{enumerate}
	\item \label{itm:step1} The primary cluster analysis and method selection step: Using the original dataset or a part of it (in the following called ``discovery data''), select a \textit{single} clustering method (where the ``method'' includes not only the choice of clustering algorithm, but also parameters such as the number of clusters and diverse pre/postprocessing steps), for example, via its performance with respect to internal/external validation indices.
	\item \label{itm:step2} The validation step: Validate important aspects of the clustering resulting from this method on another dataset or the rest of the original dataset (in the following denoted by ``validation data''). The validation data must be completely hidden from the method selection process of Step~\ref{itm:step1} -- analogously to the evaluation of supervised classifiers, where the selected model (including the chosen parameters) must be finally evaluated using validation data that was \textit{not} used in any way for parameter tuning or model selection  (\cite{boulesteix_evaluating_2008,simon_pitfalls_2003}).
\end{enumerate} 

The ``important aspects'' of the clustering that should be checked in Step~\ref{itm:step2} usually depend on the research question and the field of application. Consider the example of clustering patients with a certain disease (e.g., breast cancer or glioblastoma), based on expression levels of disease-related genes, for the purpose of finding subtypes of that disease (see for example \textcite{sorlie_repeated_2003, verhaak_integrated_2010}). In this context, the following properties might be relevant aspects of the clustering: 
\begin{itemize}
	\item Suppose that Step~\ref{itm:step1} has resulted in two clusters. One cluster is much larger than the other, with about 80\% percent of the patients in this cluster. We might be interested in whether this pattern of one large cluster and one smaller cluster can be replicated in Step~\ref{itm:step2}.
	\item Assume we have found that the clustering chosen in Step~\ref{itm:step1} is related to survival time, i.e., the patients' survival times differ depending on which cluster they belong to. Can this finding be replicated in Step~\ref{itm:step2} for patients in the validation data?
\end{itemize} 
In the literature, the term ``cluster validation'' is sometimes used to refer to the use of validation techniques as a tool to compare different clusterings and select the most appropriate. This use of terminology would place validation within Step~\ref{itm:step1}. In our framework, clusterings that were selected in this way are still to be validated in the sense of Step~\ref{itm:step2}.
%, preferably by validation techniques that were not optimised in Step~\ref{itm:step1} in order to avoid selection bias. %commented out by Theresa 

There is some work on the so-called \textit{benchmarking} of clustering methods (\cite{van_mechelen_benchmarking_2018,zimmermann_method_2020}). This is different from our approach. A benchmarking study is a systematic comparison of different clustering \textit{methods} on a class of data distributions or datasets.
% (as opposed to the evaluation of a concrete clustering \textit{result}). 
%  Such a study is supposed to provide researchers with guidance regarding the choice of a method in a wide range of applications.
In contrast, our two-step validation framework is not intended to provide general guidance on the quality of clustering methods, but to be directly used by a researcher performing clustering within a specific application, regardless of the choice of the clustering method. 
% and the distribution that actually gave rise to the data, namely here 
% on two datasets---the  ``discovery'' and the ``validation'' dataset. In Step~\ref{itm:step1}, the researcher can try different methods to select a final ``good'' solution, but the selection process does not have to follow a rigorous formal prescription, which would be required in benchmarking. This is because the result is not meant to be generalised to different distributions, and only meant to be informative for the chosen  specific clustering.

This paper is structured as follows: In Section \ref{sec:relatedwork} we give an overview of the different uses of the term ``validation'' and perspectives on validity found in the clustering literature. We discuss concepts related to our framework, such as stability analysis. We then present our validation framework  in detail in Section \ref{sec:framework} and demonstrate in an exemplary manner how clustering studies from the applied literature can be sorted into the framework. Section \ref{sec:discussion} contains a final discussion.

\section{Different perspectives on ``validity'' in cluster analysis}
\label{sec:relatedwork}

In order to explain what ``validation'' of a clustering result on validation data means and how it relates to methods commonly denoted as ``cluster validation'' in the literature, we first discuss the concept of ``cluster validity''. 
% To define ``validity'', or even to clarify which part of the clustering procedure it refers to, is not easy. 
Most importantly, it can refer to a clustering \textit{method}, or to a clustering \textit{result} on a specific dataset, and this distinction has implications for how validation should be done.

By reviewing the literature, we identified five approaches that address the validity of a clustering method or clustering result:  1) the comparison of \lq\lq true'' cluster labels with inferred clusters, 2) internal and external validity indices, 3) stability analyses, 4) hypothesis tests assessing the presence of a cluster structure, and 5) visual validation. These five approaches are briefly reviewed in the following subsections. 

\subsection{Recovery of ``true'' clusters and analogies to the validity of supervised classification models}
\label{subsec:trueclusters}
In this perspective, a clustering of a dataset is ``valid'' if it corresponds to the ``true'' cluster structure in the data. Correspondingly, a clustering method is called ``valid'' if it can recover the ``true'' clusters in the data (\cite{milligan_methodology_1987,breckenridge_replicating_1989}). A related view is presented in the paper of \textcite{dougherty_validation_2007}, which shows a connection to the term ``validity'' in the context of supervised classification. For supervised classification models, the validation of a classifier relates to estimating the \textit{prediction error} on a test set, i.e., how well the classifier can predict the known ``true'' labels of the instances in the test set. \textcite{dougherty_validation_2007} demonstrate that this approach can be transferred to cluster analysis. One can draw an analogy between classifiers and cluster operators (a cluster operator is a function that maps a set of data points to a partition of the points). The validity of a cluster operator with respect to a data distribution $\Psi$ may be determined by estimating the clustering error that quantifies how well the operator predicts the ``true'' cluster labels of data sampled from $\Psi$.

However, estimating the prediction error of the cluster operator requires datasets with \textit{known} cluster labels. The idea of \textcite{dougherty_validation_2007} thus mainly makes sense in the context of benchmark studies comparing clustering methods using simulated data with known ``true'' cluster labels. The ability of the methods to recover the true clusters may then be used as a performance criterion.  Yet, in practice cluster analysis is usually applied to real datasets for which the ``true'' cluster labels of the data points are unknown. Note that even in the rare case of a cluster analysis performed on a dataset with given ``true'' cluster labels, these may not be unique, and there might be other equally legitimate cluster structures in the data, which can be even more interesting and useful as a result of the analysis than the one previously known (see \cite{hennig_what_2015,farber_using_2010}). 

In summary, when validating a clustering on validation data, we cannot directly mimic the validation step used in supervised classification. We  must therefore look for other options to evaluate a clustering result or a clustering method on new data. 

\subsection{Internal and external validation} 
\label{subsec:internextern}
In the absence of ``true'' cluster labels,  assessing ``cluster validity'' often uses so-called internal indices or external information---leading to the terms \lq\lq internal validation'' and \lq\lq external validation'', respectively.  
\begin{itemize}
	\item \textbf{Internal validation} uses only the data that was used for clustering. Typically, internal validation consists of calculating an index that is supposed to measure how well the clustering fits the data (\cite{halkidi_method-independent_2015}). Such indices often exploit the proximity structure of the data, e.g.,\ by measuring the homogeneity and/or the separation of the clusters. Many indices are ``composite'' in the sense that they evaluate the overall quality of the clustering by taking multiple internal properties (often both homogeneity and separation) into account. Examples are the Average Silhouette Width index (\cite{kaufman_finding_2009}) and the Cali\'{n}ski-Harabasz index (\cite{calinski_dendrite_1974}). These indices combine measurements of the homogeneity and the separation of a clustering into a single value, in order to balance a small within-cluster heterogeneity and a large between-clusters heterogeneity. There are also indices that measure only isolated aspects of a clustering (e.g.,\ only the homogeneity or only the separation of the clusters), see \textcite{akhanli_comparing_2020}. Acknowledging that different aspects of clustering quality may be of varying importance in different applications, \textcite{akhanli_comparing_2020} recently introduced a method to aggregate indices measuring isolated aspects of a clustering into a composite index which is tailored to the specific clustering aim of the researcher.
	\item \textbf{External validation} makes use of additional (external) information that was \textit{not} used for clustering. For example, when clustering a cancer gene expression dataset, one may use the survival time of patients to determine whether the clustering of patients based on gene expression can predict survival. The term ``external validation'' also encompasses the recovery of previously known ``true labels'' as presented in Section \ref{subsec:trueclusters}: If a certain partition of the data is indeed previously known, then it can be compared to the calculated clustering, e.g., with the Adjusted Rand Index (ARI) (\cite{rand_objective_1971,hubert_comparing_1985}) or the FM index by \textcite{fowlkes_method_1983}. See \textcite{meila_criteria_2015} and \textcite{albatineh_similarity_2006} for overviews of partition similarity indices. 
\end{itemize}
Internal and external validation indices are very popular and can appear in our validation framework in Step~\ref{itm:step1} (for method selection) as well as in Step~\ref{itm:step2} (validation). In Step~\ref{itm:step1}, internal and external validation indices can be used to compare multiple clusterings by calculating an internal/external validation index (or several indices) for each clustering using the discovery data. One can then choose the best clustering approach, i.e., the approach that performs best with respect to the indices. This process is sometimes called \textit{relative validation} (\cite{jain_algorithms_1988,halkidi_method-independent_2015}). In the validation Step~\ref{itm:step2}, internal or external validation may be performed to assess whether similar internal or external index values are achieved using the validation data as previously obtained using the discovery data (see Section \ref{sec:framework}).\\
When using internal validation for method selection in Step \ref{itm:step1}, it may be advisable to use a composite index that balances multiple internal properties. For example, when the choice of method includes the number of clusters, then an index measuring \textit{only} the homogeneity of the clusters will tend towards choosing a high number of clusters, while an index measuring \textit{only} the separation of the clusters will tend towards choosing a low number of clusters. Usually, a compromise between such conflicting aims is desired. However, for the validation Step \ref{itm:step2}, one does not necessarily need composite indices: Comparing isolated aspects between the clusterings on the discovery and the validation data is appropriate for assessing in which exact internal properties the clusterings differ.

%Theresa commented out the following section (written by Theresa), because we do not demonstrate the data dredging effect with a data example and therefore lack ``proof'' for this claim. Also, it's already mentioned in the introduction. 
%One may wonder why the validation step~\ref{itm:step2} is necessary, if  internal or external validation was already performed in Step~\ref{itm:step1}. The problem is that, as briefly mentioned in the introduction, the interpretation of internal and external validation may be misleading when these steps are performed for several clustering methods successively and combined with selective reporting. More precisely, when a lot of different clustering approaches (including different numbers of clusters, different parameter values, etc.) are tried successively, it is likely that at least one of them will perform satisfactorily in internal or external validation just by the sheer accumulation of results. If the researcher primarily reports this one and puts the others in the file drawer, the resulting picture is over-optimistic. In other words, we suspect that ``researcher degrees of freedom'' (\cite{simmons_false-positive_2011}) when performing clustering may lead to overoptimistic statements as a result of overfitting the discovery set. This is what makes the validation step~\ref{itm:step2} so important. 
%We will demonstrate this with a data example in Section \ref{sec:dataexample} below. 

\subsection{Stability}
\label{subsec:stability} 
Many authors consider \textit{stability} to be a crucial aspect of cluster validity.
The idea of stability analysis is as follows: a good clustering method should yield similar partitions when applied to multiple datasets drawn from the same data distribution  (\cite{ben-david_sober_2006}). In this spirit, a specific clustering of a single real dataset may be considered as validated if the clusterings obtained from datasets generated from the same data distribution are similar. There are several methods of generating multiple datasets to emulate the data distribution of the dataset to be analysed, e.g., by drawing subsamples from the original dataset (\cite{hennig_cluster-wise_2007}).\\
In stability analysis, it is a crucial issue to define what is meant by ``obtaining similar partitions on multiple datasets''. This problem is handled differently by different authors. In this section we keep the discussion more general, and defer the concrete description of a stability procedure to Section \ref{subsec:valframework}, once important notation has been defined.\\ 
Stability analysis dates back to \textcite{mcintyre_nearest-centroid_1980}, \textcite{morey_comparison_1983}, and \textcite{breckenridge_replicating_1989}. At the time, this approach was also called ``replication analysis'' and referred to the replicability of a clustering result on a validation dataset. To generate the validation dataset, the original data is split into two halves (by splitting along the objects to be clustered). This is followed by assessing whether the clustering obtained on the first half can be replicated on the second half. The idea of splitting the data in such a way is the basis of our validation framework, and such stability analyses will indeed be a special case of the broader validation framework presented in Section~\ref{sec:framework}.

%More precisely, points \ref{itm:valint} and \ref{itm:valext} are usually not considered in stability studies.
In the decades that followed, however, the focus of stability analysis shifted away from this concept and more towards \textit{method or model selection}. Like other validation techniques, stability analyses are used in Step~\ref{itm:step1} as a basis for the selection of a suitable clustering method and its parameters, such as the number of clusters (\cite{levine_resampling_2001,dudoit_prediction-based_2002,ben-hur_stability_2002,monti_consensus_2003,lange_stability-based_2004,tibshirani_cluster_2005,bertrand_loevingers_2006,dolnicar_evaluation_2010,wang_consistent_2010,fang_selection_2012}). In these approaches, stability analysis selects the clustering method that is most stable over multiple subsamples. The subsamples are drawn without replacement or in a cross-validation manner, or are bootstrap samples drawn with replacement from the data. For example, different numbers of clusters $k$ can be considered in turn, and the $k$ that leads to the most stable clustering, or the smallest $k$ that exceeds a stability threshold, can be chosen.

%It would then be inappropriate to use the same stability results in the validation step \ref{itm:step2}. %commented out by Theresa
When splitting the dataset multiple times to determine the stability of a clustering method or parameter, eventually information from the whole dataset enters the method selection process. In contrast, as already mentioned above, we argue for (additionally) putting aside a validation dataset that is only used \textit{after} the method selection. Even if a clustering is chosen by stability analysis, it is \textit{not} guaranteed that this clustering can be validated on a validation dataset.  %although it is conceivable, with large computational effort, to use nested subsampling schemes to investigate the stability of a clustering selected by optimising stability.
%Theresa changed this to:  
Using a validation dataset in this manner may be considered as a simple version of a nested resampling scheme: The inner resampling loop corresponds to the method selection via the stability procedure on the discovery set, and the decision is then evaluated on the outer loop (the validation dataset). In principle, it is conceivable (with potentially large computational effort) to perform multiple resamplings in the outer loop, i.e., to split the dataset multiple times into discovery and validation datasets. However, we will not pursue this further here.

As stability refers to finding similar clusterings on similar datasets, a key issue for stability measurement is how different clusterings on different datasets can be compared. Sometimes clusterings on subsamples are compared to clusterings on the whole dataset (e.g., \cite{levine_resampling_2001,bertrand_loevingers_2006,hennig_cluster-wise_2007,dolnicar_evaluation_2010}). This can be done using external validation indices such as the ARI (see Section \ref{subsec:internextern}) on the subsample only. Some schemes require the comparison of clusterings on samples that may have small or no overlap, usually when one compares clusterings on different, often disjunct, subsamples of the original dataset (e.g., \cite{dudoit_prediction-based_2002,lange_stability-based_2004,tibshirani_cluster_2005,wang_consistent_2010,fang_selection_2012}). This 
%is usually also done by partly tailor-made external validation indices, but it %commented out by Theresa
requires a supervised classification step for classifying observations of one sample to the clusters of the other sample. This can also be required in our framework when comparing a clustering on the discovery data with a clustering on the validation data.
Most of the original papers only mention assignment to the closest cluster mean, but this is not appropriate for all clustering methods, and alternatives are listed in \textcite{akhanli_comparing_2020}. 

How does stability relate to the other validation approaches discussed in Sections \ref{subsec:trueclusters} and \ref{subsec:internextern}? Stability is certainly not equal to ``validity'' in the sense of finding the ``true'' clusters -- a clustering may be stable even though it does not correspond 
% a) to a ``natural'' cluster structure in the data (the case when the data is actually homogeneous or b) 
to the ``true'' cluster structure in case this exists, or even though the underlying data is homogeneous. % last part of the sentence added by Theresa
This has been demonstrated extensively both theoretically (\cite{ben-david_sober_2006,von_luxburg_clustering_2010}), as well as empirically by using simulated data (\cite{krieger_cautionary_1999,brun_model-based_2007,steinley_choosing_2011,senbabaoglu_critical_2014}). The fact that assigning all observations into the same cluster will always be stable illustrates that although a good amount of stability may be a reasonable requirement for a valid clustering, optimising it may not lead to the ``best'' clustering.  % However, a stable clustering found by the optimisation procedure might be interesting in its own right (\cite{hennig_cluster-wise_2007}). 

Stability indices could be considered to be internal validation indices that are used for relative validation and method selection, see Section \ref{subsec:internextern}. For example, \textcite{handl_computational_2005} classify stability indices under ``internal measures'', while noting their differences in comparison with ``classical'' internal validation indices (such as the Average Silhouette Width by \textcite{kaufman_finding_2009}). The latter are often based on proximity-related properties such as compactness or separation, and do not require the repeated application of clustering algorithms.
%, which implies that apart from the data themselves also information about the clustering algorithm is required that led to the clustering. %commented out by Theresa 

Stability analysis can also be combined with classical internal validation indices by checking whether internal validation indices have similar values for multiple clusterings calculated on subsamples of the data (\cite{jain_bootstrap_1987}). This idea will also be featured in our framework in Section \ref{sec:framework}.

\subsection{Null models and hypothesis testing}
\label{subsec:testing}
Sometimes the validity of a clustering is judged by whether or not it rejects the null hypothesis of a statistical test. For example, \textcite{dubes_cluster_1993} writes that a cluster structure is valid ``if it cannot reasonably be assumed to have occurred by chance or to be an artifact of a clustering algorithm. Validation is accomplished by carefully applying statistical methods and testing hypotheses.'' Here the question arises of what exactly ``by chance'' means, or put differently: What is the null hypothesis in such a test?\\ 
Statistical tests can appear in external validation (Section \ref{subsec:internextern}) when testing whether the clustering is associated with an external variable. For example, one may use a logrank test for testing whether a clustering of patients based on gene expression is associated with survival time. The null model is the usual null hypothesis of the respective test, i.e., the assumption of no association.\\
In another category of tests the null model consists of homogeneous data without inherent cluster structure (i.e., the null hypothesis is that the clustering result is actually obtained from a data distribution that models homogeneous, non-clustered data). Sometimes the random label null hypothesis is also used, where clustering results are evaluated with respect to what would be expected under ``random'' cluster label assignment. We will discuss tests based on the homogeneous data null hypothesis and the random label null hypothesis in the remainder of this section. More comprehensive overviews of the topic of hypothesis testing for cluster analysis can be found in \textcite{jain_algorithms_1988}, \textcite{bock_probabilistic_1996}, \textcite{gordon_cluster_1998}, and \textcite{huang_significance_2015}.

\subsubsection{Null models based on homogeneous data distributions}
\label{subsubsec:testhomogen}
Tests of homogeneous data null models address the following question: Is the clustering generated by the clustering method $M$ from the given dataset $D$ ``better'' than clusterings generated by the same method $M$ on homogeneous datasets? The general procedure for such tests is as follows (cf.\ \cite{huang_significance_2015}):
\begin{enumerate}[label=T\arabic*)]
	\item\label{itm:t1} Determine a null model $H_0$ that formalises  homogeneity.
	\item Choose a statistic $T$ that measures the quality of the clustering. This could, for example, be an internal index or stability index (see below). Apply the clustering method $M$ to $D$ and calculate $T$ for this clustering, resulting in the value $t$.
	\item\label{itm:t3} Compute $P_T$, the distribution of $T$ under $H_0$, usually by Monte Carlo simulations, i.e., by simulating $M$ datasets from $H_0$. For each simulated dataset $D^*_l$, $l = 1,\ldots,M$, apply the clustering method $M$ to $D^*_l$ and compute the value $t^*_l$ of $T$.
	\item\label{itm:t4} Calculate a $p$-value for the given dataset $D$:
	\begin{equation*}
	\hat p = \frac{\lvert \{l: t_l^* \geq ~ t \}\rvert +1}{M+1},
	\end{equation*}
	assuming here that higher values of $T$ indicate a stronger cluster structure.
\end{enumerate}
The test statistic $T$ often connects such tests to the indices discussed in
Sections \ref{subsec:internextern} and \ref{subsec:stability}.
$T$ can be chosen as an internal validation index to check whether the index value for the dataset $D$ of interest is ``significantly better'' than index values expected for clusterings on homogeneous data (\cite{dubes_cluster_1993,gordon_cluster_1998, halkidi_cluster_2002, hennig_flexible_2015}). Alternatively, $T$ may be chosen as a stability value, to evaluate whether the method $M$ yields more stable results on the given data $D$ than on homogeneous datasets (\cite{smith_stability_1980, dudoit_prediction-based_2002, bertrand_loevingers_2006}). This can be computationally intensive as the stability procedure (with multiple resamplings of the data) must be repeated for each simulated dataset $D^*_l$.\\
Sometimes the test procedure is also used for method selection, as in Step~\ref{itm:step1} of our proposed two-step procedure, mostly for determining the number of clusters (see for example, \cite{tibshirani_estimating_2001, dudoit_prediction-based_2002, hennig_flexible_2015, john_m3c_2020}). In this case, the test has to be repeated for different numbers of clusters $k$ in a certain range and the estimate for the correct number can be taken, e.g., as the $k$ that yields the smallest $p$-value in \ref{itm:t4}. However, due to the multiple testing problem, such $p$-values should not be interpreted in terms of classical error probabilities.\\
Regarding the choice of the null model for homogeneous data in \ref{itm:t1}, there are several options. For example, the usual choices for Euclidean data are the uniform distribution on a suitable set or a (single) multivariate Gaussian distribution (\cite{gordon_cluster_1998}). Yet, the class of conceivable null models can be quite large, and one may wish to incorporate more structure found in the dataset to be analysed into the null model. Otherwise, there is a danger to reject the null model not because of the existence of a meaningful clustering, but rather because the null model was unrealistic for other reasons such as a non-modelled dependence structure of the observations, see \textcite{hennig_flexible_2015}.\\ 
Ultimately, homogeneity tests are only reasonable to apply when the researcher actually requires the dataset to have a non-homogeneous structure. In some applications this is not the case. For example, in market segmentation the researcher may be quite content with clusters that are homogeneous, but not notably separated (\cite{dolnicar_evaluation_2010,hennig_what_2015}). A dataset used for this purpose may not reject the null hypothesis of a homogeneity test, but a clustering of this dataset can still be useful for the application at hand.

\subsubsection{Null models related to random cluster labels}
\label{subsubsec:testother}
Constructing null hypotheses is also possible based on ``random cluster labels'' or ``random partitions''.  This means that a clustering result is evaluated in comparison with a ``random'' assignment of cluster labels, treating the data as fixed.\\ 
Random partition or random label hypotheses are suited to some aspects of external validation (see Section \ref{subsec:internextern}), when comparing a clustering calculated on a dataset with an externally known partition, not derived from the data themselves. A partition similarity index, such as the Rand index or Jaccard coefficient, can then be computed. One way to incorporate a random partition hypothesis into such indices is to correct the index ``for chance'', i.e., by taking the expected value of the index under the assumption that both partitions are assigned randomly as the baseline. For example, the Adjusted Rand Index (ARI) is calculated from the Rand Index (RI) using such a correction process (see \cite{hubert_comparing_1985} for details). 

Conducting a test procedure similar to \ref{itm:t1}-\ref{itm:t4} with a suitable implementation of the random label hypothesis as the null model is also possible. Using external indices as test statistics assumes that the partitions to be compared are independent (\cite{hubert_comparison_1977}, \cite{wallace_method_1983}). This assumption is fulfilled when comparing a clustering to a partition that was given independently of the data to be clustered, but it is usually not fulfilled in stability analysis, where clusterings are derived from (at least partly) the same data (\cite{dudoit_prediction-based_2002}). 
To our knowledge, hypothesis tests based on the random label null model are hardly ever done. Internal validation indices should not be used as test statistics because it is likely that random partitions yield worse values regarding internal indices than clusterings chosen on the data. A clustering of even homogeneous data may thus look significantly better in this situation (see Chapter 4 of \textcite{jain_algorithms_1988} for details).

\subsection{Visual validation}
Cluster analysis is often exploratory without fixed predefined expectations from 
the user. Patterns in the data that qualify to be interpreted as clusters can 
have very diverse appearances. Some key characteristics of clusters, such as
being areas of high density separated by areas of lower density, are difficult 
to translate into easily computable statistics. Furthermore, many clustering 
methods rely on model assumptions and cluster concepts, the appropriateness of
which is hard to diagnose by means other than visual. 
This explains why visual 
validation is important in cluster analysis. It gives the user a
more holistic view of the structure manifested in a clustering, and enables them to check
whether the clustering corresponds to visually distinct patterns of the
data that do not need to be prespecified. Clusters can be declared valid based
on visualisation if they correspond to clearly visible patterns in the data,
or in some cases if the assumptions required for the chosen clustering method
look valid.

Useful plots for visual cluster validation can be distinguished into:
\begin{enumerate}
\item General purpose data plots in which found clusters can be indicated by
colours or glyphs, such as scatterplots, matrix plots, principal components
biplots, multidimensional scaling, or parallel coordinates plots 
(Chapter 5 of \cite{cook_swayne_2007}). There are also projection pursuit 
approaches that generate ``interesting'' data projections, potentially showing
clustering structure, without requiring the clustering as input 
(e.g., \cite{tyler_et_al_ics_2009}).
\item Plots set up to visualise a specific clustering, which can be further
classified as:
\begin{enumerate}
\item[(a)] Plots that visualise the original data directly, such as cluster
heatmaps (\cite{wilkinson_friendly_2009,hahsler_hornik_2011}) or projections
to optimally discriminate clusters (\cite{hennig_2004}).
\item[(b)] Plots that visualise the clustering solution without representing
the original observations directly such as dendrograms, silhouette plots, and 
neighbourhood graphs (\cite{leisch_2008}).
\end{enumerate}
\end{enumerate}
Plots that visualise the original data directly can be used to assess patterns
in data space, although these plots come with either information loss by 
dimension reduction, or heavy 
reliance on aspects such as variable and observation
ordering. The advantage of plots that optimise objective functions dependent
on the clustering, such as discriminant projections or heatmaps with orderings
determined by the clustering, is that they have better chances to bring out the
data patterns corresponding to the clustering than general purpose plots. On the other hand, they may lead to an overoptimistic assessment of the validity
of the clustering, or an interpretation of spurious patterns. 
In some places (\cite{buja_et_al_eda_inference_2009,hahsler_hornik_2011}), 
plots are compared with plots generated from ``null model'' 
random homogeneous data in order to assess whether the observed structure is 
``significant'' (see Section \ref{subsubsec:testhomogen}). Validation data that is kept separate from the beginning of 
the analysis can also help, see Section \ref{subsec:valframework}.  

Some of the plots that do not represent the original observations directly
can also be valuable for cluster validation. The silhouette plot accompanies
the Average Silhouette Width index (\cite{kaufman_finding_2009}) and gives 
observation-wise information about the quality of assignment in the given
clustering; dendrograms visualise the hierarchical merging process and can
sometimes reveal issues, such as potentially meaningful clusters disappearing at
higher levels of the hierarchy.  

\section{A systematic framework for validating a clustering on a validation dataset}
\label{sec:framework}
In this section, we present a systematic framework for validating a clustering on a validation dataset that includes some approaches from the literature as special cases and revisits them more formally. While applied researchers sometimes validate their results on a validation set, the validation strategies are  not usually discussed in a systematic and comprehensive manner. Some strategies are also problematic from our point of view, for example, procedures that already use the whole dataset in the method selection Step~\ref{itm:step1}. These do not strictly set apart the validation set for later validation. Method selection executed this way is likely to be biased towards successful validation. In general, different validation strategies are scattered across different works and application fields. We aim to give a comprehensive guide for researchers, enabling them to make an informed decision about what validation strategies they may use for their purposes.\\
We first discuss what we mean by a ``validation dataset'' in Section~\ref{subsec:valdatadef}. In Section~\ref{subsec:valframework} we give an overview of properties of a clustering result that may be validated on the validation set (this will also demonstrate the relation to the classical validation procedures discussed in Section \ref{sec:relatedwork}). In Section \ref{subsec:valperform} we explain how validation of each of these properties can be performed in practice. In Section \ref{subsec:replication} we define the term ``replication'' in the context of our validation framework. In Section \ref{subsec:successfulvalid} we discuss how to judge whether ``successful'' validation or replication has been achieved. Finally, in Section~\ref{subsec:appliedlit}, we give some examples from the applied literature, and demonstrate how these examples fit into our framework. 

\subsection{What does the validation data look like?}
\label{subsec:valdatadef}
The term ``validation dataset'' can refer to a dataset composed of independently collected data (e.g., collected by other researchers or in a different laboratory) which is similar enough to the original data for cluster evaluation to be possible. 
In practice, however, genuinely independent data is often not available. In this case, one might split a single dataset into a discovery and a validation set.\\
In the following, we always consider a fixed data distribution $\Phi$ from which both the original data and the validation data are drawn; this is an idealised assumption, particularly if the validation data is an independently collected dataset. The validation of an aspect of the clustering should always be considered \textit{with respect to this data distribution}.\\
Data sampled from $\Phi$ can either be object by variable data or object by object (dis)similarity data (\cite{van_mechelen_benchmarking_2018}). Here, ``objects'' denotes the entities which are to be clustered.\\
To further clarify what properties a validation dataset must fulfill, it is important to distinguish between what we denote as \textit{inferential} and \textit{descriptive} clustering:
\begin{itemize}
	\item {\bf Inferential clustering:} The objects being clustered form a sample drawn from an underlying population for which inference is of interest, rather than making statements about the specific objects in the original dataset. 
% They are not interesting in themselves but only as representatives of the population. Consequently, the results of the clustering analysis will not include any statement such as \lq\lq objects 1, 5 and 99 form a cluster'', because it does not matter which of the objects cluster together. Clustering analyses in this setting essentially aim at generating knowledge about the whole population through the analysis of a (hopefully) representative sample, hence the term \lq\lq inferential''.
	\item {\bf Descriptive clustering:} The data form a fixed set of entities of specific interest, and statements such as \lq\lq objects 1, 5, and 99 form a cluster'' are of interest.
\end{itemize}
As an example of the difference between inferential and descriptive clustering, consider an $n\times p$ dataset including the expression levels (continuous values) of $p$ genes for $n$ patients suffering from a particular disease. It may be of interest to perform clustering analyses of the patients to see if there are subpopulations of patients with systematically different gene expressions. This would be {\it inferential clustering}. It may also be of interest to perform clustering of the (fixed set of) genes to see if there are groups of specific genes that behave similarly, which might suggest a similar function or involvement in a common pathway. This is an example of {\it descriptive clustering}.\\
% In summary, the $n\times p$ gene expression dataset can be considered as a object by variable dataset in two ways: One, with the patients as the objects and the genes as the variables, and two (in transposed form) with the genes as the objects and the patients as the variables.\\  
A validation dataset for inferential or descriptive clustering (either obtained by splitting the original dataset or by collecting a separate dataset) must be as follows:
\begin{itemize}
	\item \textit{Validation data for inferential clustering}: Here the validation data consists of more objects to cluster. Consequently, if the original dataset is split into discovery and validation sets, the split is performed along the objects.\\
	If the dataset is given in \textit{object by variable form}, this means that the variables in  discovery and validation data are the same.
% In the example of the $n \times p$ gene expression dataset, to split this into discovery and validation sets we divide the patients into two groups.\\
	If the dataset is given in \textit{object by object (dis)similarity form}, the objects can be split into two disjoint sets, thus yielding two smaller (dis)similarity matrices (one representing the discovery data, the other the validation data) from the original (dis)similarity matrix.\\
	In some cases, it could make sense to choose a validation dataset that consists of more variables, rather than more objects. For example, \textcite{smolkin_cluster_2003} use such an approach for an $n \times p$ gene expression dataset to check whether patient clusters are stable when using different genes for the clustering. This implies that the clustering is meant to be stable over different variables, which in reality may or may not be the case.
% In order for such a validation procedure to work, the genes (variables) in the discovery set need to characterise the clusters in the same ways as the genes in the validation set do. If subtypes of an illness are characterized by just a few important genes, and these genes or some of them are not present either in the discovery or validation set, then the validation procedure will likely fail. 
In the remainder of this paper, we will not consider this procedure further, but instead always assume that a validation dataset for inferential clustering consists of more objects. 
	\item \textit{Validation data for descriptive clustering}: The validation data does \textit{not} consist of more objects to cluster, because the set of objects to be clustered is fixed.\\
	If the data is given in \textit{object by variable form}, the validation dataset consists of more variables. Correspondingly, a split of the original dataset into discovery and validation sets must be performed along the variables. For the example of the $n \times p$ gene expression dataset, now the patients are considered as the ``variables'', and the genes as the objects. Thus, if the data is split into discovery and validation set, the patients are split into two groups, as in inferential clustering.\\
	If the dataset is given in \textit{object by object (dis)similarity form}, then a validation dataset is a (dis)similarity matrix of the same objects, but with their (dis)similarities derived from another source (e.g., based on different underlying variables). In this case, if only a single dataset is available, it is not possible to split the given proximity data into discovery and validation data, although it may be possible to split the underlying variables. %brought back in by CH.
Also, even in descriptive clustering, the stability of the clustering pattern, as to be investigated by splitting the objects to be clustered, may sometimes be of interest, but we do not pursue this here. %brought back in by CH.
\end{itemize}
If separately collected data is not available and the dataset must be split into discovery and validation sets, we believe that a 50/50 split ratio makes sense in most cases: In our opinion, validation strategies often require the number of data points in the validation set to not be too small when trying to validate certain properties obtained from the clustering on the discovery set. A similar argument has been made in the context of stability analysis (\cite{lange_stability-based_2004}).
%\cite{lange_stability-based_2004} argue: ``[T]he data sets should have (approximately) equal size so that an algorithm can find similar structure in both data sets. If there are too few samples in one of the two sets, the group structure might no longer be visible for a clustering algorithm.''\\

\subsection{Which properties of a clustering may be validated?}
\label{subsec:valframework}

What are the aspects that can be validated in Step~\ref{itm:step2}?
We sort these properties into four different categories:

\begin{itemize}
	\item[\namedlabel{itm:valint}{(Int)}] Internal properties of the clusters (that turn up after clustering with the discovery data), for example: 
	\begin{itemize}
		\item descriptive measures of the clusters such as the values of the cluster centroids or the relative sizes of the clusters,
		\item the value of an internal validation index calculated for the clustering result, %(cf.\ ``internal validation'' in Section \ref{subsec:internextern}),
		\item subsets of variables that characterise the clusters.
	\end{itemize} 
	\item[\namedlabel{itm:valext}{(Ext)}] Associations of the clusters with external variables or agreement of the clustering with an externally known partition. % (cf.\ ``external validation'' in Section \ref{subsec:internextern}). 
Some examples:
	\begin{itemize}
		\item Clusters of cancer patients have different mean survival rates.
		%\item clusters of neonatals (based on differences in gut microbiome) are associated with different relative risks for asthma (Fujimura et al.\ 2016),
		\item A clustering of genes shows some agreement with known functional gene labels.  For example, a clustering may be fully compatible with known partitions of the genes into functional categories. Less restrictively, some particular genes, say genes X and Y, may be in the same cluster (as previously expected). 
	\end{itemize}
	\item[\namedlabel{itm:valvis}{(Vis)}] Characteristics that can be assessed using visualisation: Do the clusters correspond to distinctive meaningful patterns in the data? Do the clusters look how they were supposed to look like? This could refer to model assumptions for the clustering method, or {\it a priori} hypotheses or requirements by the researcher. 
	\item[\namedlabel{itm:valstab}{(Stab)}] Stability of cluster membership: Does cluster membership remain stable when the same method (algorithm, number of clusters, etc.) is applied to the validation data? %As the drawn objects for the clusterings to be compared are typically different or even disjunct (e.g., in case of discovery and validation dataset), % Theresa changed this to: 
	Since the objects in the discovery and the validation set are disjunct in the case of inferential clustering, this may involve supervised classification of objects of one dataset to clusters of the other dataset.
\end{itemize}
Most subsections of Section \ref{sec:relatedwork} correspond to a category in the above list, with the exception of Section \ref{subsec:trueclusters} (recovery of ``true'' clusters) and \ref{subsec:testing} (hypothesis testing). As shown in Section \ref{subsec:testing}, hypothesis tests are usually related to internal, external, or stability properties, so may be used in the categories \ref{itm:valint}, \ref{itm:valext}, or \ref{itm:valstab}, respectively.\\
We denote the discovery data by $D_{1}$ and the validation data by $D_{2}$. The clustering chosen on $D_1$ in Step~\ref{itm:step1} is called $C_1$. Given $C_1$, the validation dataset can be handled in two different ways:
% For \ref{itm:valint}, \ref{itm:valext} and \ref{itm:valvis} the major interest can be to validate the use of the selected \textit{clustering method} or rather the \textit{clustering result for the specific objects}.\\
\begin{itemize}
\item[(a)] The same clustering method that yielded $C_1$ (i.e., same algorithm, same number of cluster $k$, etc.) can be applied to $D_2$, yielding a clustering $C_2^{md}$ on $D_2$ (``md'' for ``method''). $C_1$ and $C_2^{md}$ can then be compared with respect to aspects \ref{itm:valint}, \ref{itm:valext} or \ref{itm:valvis}. We call this approach \textit{method-based validation}. It puts a focus on the structural similarity of the clustering results as generated by the method. 
\item[(b)] Instead of applying the clustering method again, $C_1$ can be ``transferred'' to the validation data by using a supervised classifier to predict the cluster labels of the validation set (explained in more detail below). This results in a clustering $C_2^{tf}$ on $D_2$ (``tf'' for ``transferred''). The transferred clustering can be compared to the original clustering $C_1$ with respect to aspects \ref{itm:valint}, \ref{itm:valext}, or \ref{itm:valvis}. We call this approach \textit{result-based validation}. It puts a focus on whether the specific clustering result is also sensible for the validation data.\\
We now explain what we mean by ``transferring'' the clustering. For descriptive clustering, $C_2^{tf}$ is simply $C_1$ (recall that for descriptive clustering, the objects to be clustered are the same for $D_1$ and $D_2$, and thus $C_1$ can immediately be considered to be a clustering of $D_2$). For inferential clustering,  the objects to be clustered are different in the discovery and validation sets, so some proper ``transfer'' is required. This can be done using a supervised classifier (using the labeled discovery set $(D_1, C_1)$ as ``training set'') to assign the objects in $D_2$ to the clusters in $C_1$ (\cite{lange_stability-based_2004, akhanli_comparing_2020}). For example, one can calculate the centroids of the clusters in $C_1$, and then assign each sample in $D_2$ to its nearest centroid (``nearest-centroid classifier''). As $C_2^{tf}$ is supposed to be an ``extension'' or ``transfer'' of the original clustering to the validation data, one should use a classifier that fits the assignment rule of the chosen clustering algorithm as closely as possible. Hence the nearest-centroid classifier is suitable for $k$-means, which indeed clusters points by assigning them to the nearest centroid. For suitable classifiers for other clustering algorithms, see \textcite{akhanli_comparing_2020}.
\end{itemize}
For \ref{itm:valstab} (stability of cluster membership) the clustering method necessarily has to be applied again to $D_2$. We check whether the cluster memberships resulting from applying the method to the validation data are similar to the cluster memberships resulting from transferring the original clustering to the validation data. This may be considered as a combination of method-based and result-based validation.

\subsection{Performing the validation}\label{subsec:valperform}
We now explain in general terms how to perform the validation steps with respect to the categories \ref{itm:valint}, \ref{itm:valext}, \ref{itm:valvis}, and \ref{itm:valstab} given above, see also Table \ref{tab:valprocedure}. 

In practice, many decisions are left to the user and the precise choice of indices, plots, etc.\ will depend on the context and aim of the analysis, see also Section \ref{sec:discussion}.
 
\begin{table}[h]
	\caption{Strategies for validation on validation data}
	\label{tab:valprocedure}
	\centering
		\begin{tabular}{r |c|c| } 
			\cline{2-3}
			& method-based validation & result-based validation \\ 
			\cline{2-3}
			& compare $C_1$, $C_2^{md}$ with respect to & compare $C_1$, $C_2^{tf}$  with respect to\\\ 
			\ref{itm:valint} & internal properties, & internal properties, \\
			\ref{itm:valext} & external associations, & external associations, \\
			\ref{itm:valvis} & visual properties & visual properties \\
			& \multicolumn{2}{|c|}{compare $C_2^{md}$, $C_2^{tf}$ with respect to} \\
			\ref{itm:valstab} & \multicolumn{2}{|c|}{cluster membership} \\
			\cline{2-3}
		\end{tabular}
\end{table}

\subsubsection*{Validating \ref{itm:valint}: Internal properties of the clustering}
\textbf{Procedure:} For \textit{method-based validation}, compare the internal properties of $C_2^{md}$ to those of $C_1$. For \textit{result-based validation},
%(is used e.g.\ in Kapp/Tibshirani 2007): 
compare the internal properties of $C_2^{tf}$ to those of $C_1$. For example, when using internal validation indices, calculate the index both for $C_1$ and $C_2^{md}$ (resp.\ $C_1$ and $C_2^{tf}$), and compare the two index values. 
\\
\textbf{Remarks:}  When applying result-based validation, the clusters of $C_2^{tf}$ correspond one-to-one to those of $C_1$. This makes the comparison easier.
In method-based validation, the clusters of $C_2^{md}$ are not automatically associated one-to-one with the clusters of $C_1$. Such an association is not needed when calculating internal indices that refer to a whole clustering (i.e., are not calculated cluster-wise). But one may also be interested in comparing cluster-wise characteristics such as cluster centroids, and in this case there needs to be a matching of the clusters of $C_2^{md}$ to the clusters of $C_1$, usually assuming that their number is the same. There are various methods to do this. For example, in centroid-based clustering one could match the centroids so that the sum of distances between centroids of matched clusters is minimum (\cite{mirkin_clustering_2005}).  \textcite{breckenridge_validating_2000} suggests to associate each cluster of $C_2^{md}$ to a cluster of $C_2^{tf}$ (e.g., by choosing the cluster association that maximises the sum of the intersections of the clusters). Then the one-to-one cluster association of $C_2^{tf}$ to $C_1$ can be used to assign each cluster of $C_2^{md}$ to one of $C_1$.

%For other internal properties (e.g., values of cluster centroids) it can be more difficult to see whether the clusters in $C_2^{md}$ have similar patterns as the ones in $C_1$. In order to associate each cluster of $C_2^{md}$ to one of $C_1$, it is an option to use $C_2^{tf}$ (\cite{breckenridge_validating_2000}): First, associate each cluster of $C_2^{md}$ to a cluster of $C_2^{tf}$ (e.g., by choosing the cluster association that maximises the sum of the intersections of the clusters). Then use the one-to-one cluster association of $C_2^{tf}$ to $C_1$ to assign each cluster of $C_2^{md}$ to one of $C_1$. This, however, requires calculation of $C_2^{tf}$, which might be not desired if the researcher only wants to conduct method-based validation (also recall from above that $C_2^{tf}$ is very dependent on the choice of classifier). Some alternatives are possible. For example, when the clustering is centroid-based, one could match each centroid (and thus each cluster) of $C_2^{md}$ to one of $C_1$  (\cite{mirkin_clustering_2005}). \\

%Theresa rephrased the next paragraph
Note that when applying result-based validation, $C_2^{tf}$ is necessarily similar to $C_1$ by construction, and over-optimistic interpretation needs to be avoided. One could compare observed similarities to results that would be obtained for homogeneous data generated from some null model (see Section \ref{subsubsec:testhomogen}). Such ``reference data'' could be generated for the validation data alone (e.g., by using a null model that may match certain descriptive statistics such as means, variances, or variable ranges of the validation data), but in some situations it might be insightful to also generate null model reference data for the discovery data. 

% one must be careful to not be too optimistic when interpreting the validation results. Since information from the first clustering $C_1$ is used for constructing $C_2^{tf}$, $C_2^{tf}$ is already heavily ``biased'' to be similar to $C_1$.  One should therefore not be surprised if validation appears to be better for result-based validation than for method-based validation. 
\subsubsection*{Validating \ref{itm:valext}: Associations with external variables or agreement with externally known partitions}
\textbf{Procedure:} For \textit{method-based validation}, check whether the associations of the clusters in $C_1$ with an external variable/known partition still hold for $C_2^{md}$. For \textit{result-based validation}, check whether the associations of the clusters in $C_1$ with an external variable/known partition still hold for $C_2^{tf}$.\\
\textbf{Remarks:} As for method-based validation of internal properties \ref{itm:valint}, here too it may be necessary to match the clusters of $C_2^{md}$ to those in $C_1$ and the remarks made above apply. This is not necessarily required. For example, testing whether the clusters are associated with an external variable, such as survival time, without interpreting the association of specific clusters does not require matching. \\
For result-based validation again the similarity by construction between $C_2^{tf}$ and $C_1$ can encourage overoptimism and may be assessed by null model simulation. For descriptive clustering, the partition $C_2^{tf}$ is actually equal to $C_1$. This makes certain approaches such as testing an association between cluster membership and an external variable meaningless.
%\textcolor{red}{Another possibility (see \cite{parmar_radiomic_2015}): use discovery data to train predictive model, and then evaluate the fit of this model with the validation data (using MMCE, CI, AUC,...). The two strategies of Paramar (descriptive clustering!) are either to calculate CI/AUC for each individual feature and then take the mean for the features in each cluster, or to take the medoids of the clusters (one feature per cluster) and train models with these as independent variables.}

\subsubsection*{Validating \ref{itm:valvis}: Visual patterns} 
\textbf{Procedure:} For \textit{method-based validation}, check whether associations of visual patterns with clusters discovered in $C_1$ still hold for $C_2^{md}$. For \textit{result-based validation}, check whether associations of visual patterns with clusters discovered in $C_1$ still hold for $C_2^{tf}$.\\
\textbf{Remarks:} Using the same variables for $D_1$ and $D_2$ 
as in inferential clustering, some
plots such as scatterplots or parallel coordinates plots can visualise both $C_2^{md}$ and $C_2^{tf}$ in a straightforward manner comparable to $C_1$. 
Some other plots such as principal components biplots, other linear 
projection plots such as those in \textcite{hennig_2004}, and 
multidimensional scaling require a selection of an optimal projection space 
for the
dataset to be plotted. Although this could be done on the validation data, 
for inferential clustering plotting the validation dataset on the projection 
space defined by the discovery dataset (and its clustering, if the projection 
space depends on it) allows a more direct comparison. 
For linear projection methods, this requires a standard linear projection 
given the coordinate axes determined from $D_1$. 
For multidimensional scaling, there are techniques 
to embed new observations into the projection space defined by the original
observations, e.g., \textcite{gower_1968}. For descriptive clustering, on the other 
hand, embedding the observations of $D_2$ in the space defined by $D_1$
is not informative as the points would be the same,
so here a projection space optimised for $D_2$ has to be found. 

Some other plots, such as the silhouette plot and cluster heatmaps (as long 
as observations are ordered only by a partition rather than a full
dendrogram), may benefit from matching clusters for determining their order, 
see the comments on internal validation \ref{itm:valint}. Matching the
dendrograms by which a cluster heatmap is ordered is a challenging problem and would benefit from further research. 
%We are not aware of any published work that uses silhouette plots, cluster heatmaps, or in fact any plot that is not a linear projection method with a separate validation dataset, although this is certainly conceivable and may be a promising direction of research. %Theresa commented this out because there might be some examples (see example in Table in Section \ref{subsec:appliedlit} which used heatmaps), even though the procedures in these studies are not sophisticated (e.g. they compare heatmaps, but don't care about matching the dendrogram, just match the clusters). I've added a couple of sentences in Section \ref{subsec:appliedlit}. 

The results of visual validation are subjective, and although plots 
are reproducible given both discovery and validation datasets, 
the way the researcher arrives at a validity verdict will not be reproducible. For orientation it may help, as before, 
to compare results on the
validation data with results on artificially generated 
null model data to see whether the
patterns on the actual validation data are more distinctive. Displaying the involved plots will give the reader the chance to make up their own 
mind.  

\subsubsection*{Validating \ref{itm:valstab}: Stability of cluster membership}
\textbf{Procedure:} Here one needs to compute both $C_2^{tf}$ and $C_2^{md}$. These are then compared with an index for comparing partitions. The rationale behind this is as follows:  cluster memberships in $C_1$ and $C_2^{md}$ are compared to check whether repeated application of the clustering method leads to stable cluster memberships. For descriptive clustering, $C_1$ can be compared to $C_2^{md}$ directly (here $C_1$ is equal to $C_2^{tf}$). 
For inferential clustering, $C_1$ and $C_2^{md}$ cannot be compared directly, because they are partitions of different sets of objects. Thus $C_2^{tf}$ is used as a surrogate for $C_1$ on $D_2$.\\
\textbf{Remarks:} Different choices of a partition similarity index are possible, e.g., the ARI, the Jaccard index or the FM index (see \textcite{meila_criteria_2015} and \textcite{albatineh_similarity_2006} for overviews). 
% Recall from Section \ref{subsubsec:testother} that indices adjusted under the random label null model, such as the ARI, can be used, but should usually not be interpreted in terms of statistical significance. % Repetition, killed. CH

\subsection{Link to replication studies}
\label{subsec:replication}
Choosing a suitable validation procedure is also important when \textit{replication} is the aim of the study. We consider ``validation'' to be the broader term, and ``replication'' as more specific, for which certain strategies of the validation framework can be used. Generally, ``replication'' refers to using new data to re-assess a scientific claim made in a previous publication (\cite{nosek_what_2020}). For clustering, this could be a claim related to the properties discussed in Section \ref{subsec:valframework}, e.g., ``found clusters of breast cancer patients based on gene expression are very homogeneous and characterised well by their centroids'' or ``the found clustering of the breast cancer patients based on gene expressions is associated with survival time''. In inferential clustering, such claims will be related to an underlying population from which the samples were drawn, e.g., the population of ``all breast cancer patients''. That is, the first (not necessarily easy) step in a clustering replication study consists of clarifying which hypotheses and claims are under investigation. %It would be helpful if the researchers of the original study to be replicated specify what results on replication data could be constituted as replication. %Theresa changed this to:
It would be helpful if the researchers of the original study specify the expected scope of their clustering results, possibly including the distribution(s) or population(s) for which the results are expected to hold. This can be useful for the choice of a suitable replication dataset and the interpretation of the replication results. 

However, there is not yet a consensus in the literature on how a replication study should be conducted, or even what the exact definition of such a study is. Often a distinction between ``direct'' and ``conceptual'' replication is made (\cite{crandall_scientific_2016}). \textit{Direct replication} is defined as conducting a study that directly mimics the procedures of the original study (e.g., the type of study participants, the statistical analysis, etc.), whereas \textit{conceptual replication} study assesses the same hypothesis as the original study, but the implementation of the study design and the analysis might differ. The distinction between direct and conceptual replication is mostly made with experimental studies in mind, and less for clustering aims. Still, one could try to adapt these definitions to the clustering context. In the context of our validation framework, direct replication will require a separately collected dataset as validation data. The validation data is obtained through a procedure as similar to the collection of the discovery data as possible. Then method-based validation can be used (i.e., applying the same method that was used on the discovery data). For conceptual replication, the route is less clear and goes beyond our framework.\\
Usage of the terms ``direct replication'' and ``conceptual replication'' is not universally accepted. For example, \textcite{nosek_what_2020} argue that these definitions focus too much on the study procedures and technical methods. Instead, \textcite{nosek_what_2020} shift the focus back to the research claim under investigation (and the scientific theory behind it) and define replication as ``a study for
which any outcome would be considered diagnostic evidence about a claim from prior
research.'' Practically any aspect of the validation framework in Section \ref{subsec:valframework} is conceivable for conducting such a study.\\
While the above definitions of replications differ, they all require the use of \textit{new} data. That is, splitting a single dataset into discovery and validation sets is not suitable for a replication study. However, the term ``replication'' is not always used in this sense in the literature on cluster analysis. For example, \textcite{breckenridge_replicating_1989, breckenridge_validating_2000} used the term ``replication'' for what we would call ``validating \ref{itm:valstab}'' in our framework, without requiring a new dataset. 

\subsection{When is a clustering successfully validated?}
\label{subsec:successfulvalid}
What does a ``successful'' validation or replication mean?
Due to random variation, we will hardly ever achieve the exact same results on discovery and validation data.
The  problem of defining ``successful'' validation or replication does not only arise in cluster analysis, but generally in most validation or replication studies. The discussion is ongoing in the field of methodological research on replication studies (cf.\ Section \ref{subsec:replication}), mostly in the context of hypothesis tests and effect estimates. For example, \textcite{hedges_statistics_2019} and \textcite{held_new_2020} argue that, when trying to replicate a hypothesis test (that was significant on the original data), it is not enough to check whether the test on the replication data is significant again. Actually, the binary distinction between significance and insignificance may not be helpful, e.g., when comparing $p$-values of 0.04 and 0.06. Rather, we should also check whether the effect estimate in the replication study provides evidence for the claim about the effect in the original study.\\
Some clustering validation aspects are connected to significance tests,  particularly testing  for external associations in \ref{itm:valext} and testing whether a meaningful clustering pattern can be distinguished from a null model for homogeneous data. The latter may be of interest in regards to internal validation indices as well as stability and visualisation. The same caveats as regarding general replication of test results apply. 

There are also aspects of clustering validation that fit less well into the framework of hypothesis tests and effect estimates. 
%Particularly, whereas in some clustering problems the question whether there is any clustering structure at all in the data is of interest (e.g., \cite{hennig_flexible_2015}), in many other problems it is rather obvious that the data are not homogeneous, and the interest is in validating the specific found clustering structure as expressed in measurements of internal or external validity, stability, or visual displays. In such cases, the rejection of a null model formalising homogeneity, independence of an external variable, or even random partitions is not very informative. %changed by Theresa to: 
For example, while in some clustering problems the question of whether there is any clustering structure at all in the data is of interest (e.g., \cite{hennig_flexible_2015}), in many other problems it is clear that the data are not homogeneous. In such cases, the interest is in validating the specific found clustering structure as expressed in measurements of internal or external validity, stability, or visual displays, and the rejection of a null model formalising homogeneity is not informative.

Considering differences between (internal or external) validity measurements on discovery and validation data, or considering an index value for stability between discovery and 
validation sets, could in principle also be framed as a testing problem of a null hypothesis formalising some kind of 
equality of structure. To our knowledge, this has not been performed yet and is 
left as a potential direction of future research.

Regarding indices measuring some aspect of cluster quality for the discovery data, it can be expected that validation data results will not be quite as good due to selection bias originating from basing selection of the final clustering on results of the discovery data (see Section \ref{sec:intro}). Observing slightly worse values on the validation data is to be expected and does not necessarily mean that the validation has failed. However, if the results are severely worse, then this may indicate problematic overoptimism on the discovery data.  

In the absence of binary decision rules (which can be seen as 
over-simplification), we must acknowledge that the question ``is validation successful?'' cannot simply be answered with ``yes'' or ``no''. The validation dataset may deliver high or low agreement regarding various aspects (internal and external validity, stability, visual aspects) with what was found on the discovery data---where the clustering on the discovery data may already have been assessed as a weaker or stronger clustering in Step \ref{itm:step1}. For example, regarding an internal index, such as the Average Silhouette Width, it is of interest both whether the value is reasonably high on the discovery dataset alone, and whether the validation dataset supports whatever value was found on the discovery data. Guidelines or thresholds for interpreting index values are rarely given and in fact mostly arbitrary, so the researcher must rely on their understanding of the index, experience, and judgment; ultimately precise results allow the reader to make up their own mind, and the situation is not much different for quantitative indices from the obviously subjective visual validation. The researcher has the option to specify a threshold value for replication based on the results of the discovery data before doing computations on the validation data to test their own expectations. However, when publishing results there is no way for the reader to verify whether the threshold was indeed specified independently of the validation data.

\subsection{Examples from the applied literature}
\label{subsec:appliedlit} 
In this section, we will review application studies that conducted cluster analysis on a discovery set and then validated the results with a validation set. Our aim is to demonstrate how these studies fit into the framework outlined above. We start by giving a short historical overview and then present some exemplary studies in Table \ref{tab:examples}. \\
The appearance of clustering studies that used a discovery and a validation set dates back to at least the 1960s.
% (the decade in which the general topic of cluster analysis started to gain traction, see \textcite{aldenderfer_cluster_1984} for a historical overview). 
Likely, one of the first clustering studies that used a validation set was \textcite{goldstein_multivariate_1969} who clustered patients with alcohol use disorder. In our terms, they performed method-based validation with respect to internal properties. 
%  based on personality and mental health related variables. The dataset was randomly split into discovery set $D_1$ and validation set $D_2$, and the clustering method was applied both to $D_1$ and $D_2$ (i.e.\ they performed method-based validation in our terms). The authors then compared the mean profiles (centroids) of the clusters of $C_1$ and $C_2^{md}$ -- in terms of our framework, the authors validated internal properties \ref{itm:valint}.\\
Soon afterwards, the idea of validation with respect to stability of cluster membership, i.e., \ref{itm:valstab}, was developed. An early implementation of this idea was conducted by \textcite{rogers_use_1973} who clustered college freshwomen based on personality features and used discriminant analysis as the classifier to derive $C_2^{tf}$. The principle of \ref{itm:valstab} was then presented more systematically by \textcite{mcintyre_nearest-centroid_1980} and \textcite{breckenridge_replicating_1989}.\\
Nowadays applied cluster analysis studies are so numerous that it is impossible to list every cluster study that used a validation set. We therefore give only a few examples in Table \ref{tab:examples} below. Some of these studies used multiple aspects of the validation framework, but for the sake of illustration, we only present one validation type per study. We did not find an example for result-based validation of \ref{itm:valvis}. In general, there appear to be few studies which performed validation of visual properties on a validation dataset in a thorough manner, and we believe future studies would benefit from considering the procedures for \ref{itm:valvis}, outlined above.

\afterpage{
\newgeometry{margin = 2cm}
\begin{landscape}
	
	\begin{table}[t!]
		\footnotesize
		\caption{Study examples for each validation type}
		\label{tab:examples}
		%\resizebox{\textwidth}{!}{%
		\begin{tabular}{lll}
			\toprule
			Type of validation & Study reference & Description\\
			\midrule
			(Int) result-based & \begin{tabular}[t]{@{}l@{}}\textcite{kapp_discovery_2006},\\\textcite{kapp_are_2007},\\
				cancer gene expression\end{tabular} & \begin{tabular}[t]{@{}l@{}} Inferential clustering of breast cancer patients. \\
				Validation data was obtained partially by splitting datasets, \\ partially by using independently collected samples.\\
				Data clustered with hierarchical clustering. \\
				$C_2^{tf}$ is derived via nearest-centroid classification.\\
				$C_2^{tf}$ is evaluated with an internal validation index (``in-group proportion'' IGP).\\
				However, the IGP was not applied to $C_1$.  
			\end{tabular} \\ \midrule
%			(Int) method-based &   \begin{tabular}[t]{@{}l@{}}\textcite{fransen_posteriori_2014},\\
%				dietary pattern analysis\end{tabular} & \begin{tabular}[t]{@{}l@{}}Inferential clustering of persons.\\ Persons randomly split into discovery and validation set.\\ Data clustered with $k$-means.\\ Clusters of $C_1$ and $C_2^{md}$ were matched manually.\\ Compared mean food intakes (centroids) between $C_1$ and $C_2^{md}$.\end{tabular}\\ \midrule
			(Int) method-based &   \begin{tabular}[t]{@{}l@{}}\textcite{bourdeaudhuij_family_1998},\\
			health behaviour of adolescents\end{tabular} & \begin{tabular}[t]{@{}l@{}}Inferential clustering of adolescents.\\ Two separately collected datasets.\\ Data clustered with hierarchical clustering.\\ Clusters of $C_1$ and $C_2^{md}$ were matched manually.\\ Compared means of health behaviour variables (centroids) between $C_1$ and $C_2^{md}$.\end{tabular}\\ \midrule
			(Ext) result-based & \begin{tabular}[t]{@{}l@{}}\textcite{curtis_genomic_2012}, \\ cancer gene expression \end{tabular} & \begin{tabular}[t]{@{}l@{}} Inferential clustering of breast cancer patients. \\
				Validation data is a second cohort.\\
				Data clustered with iCluster (\cite{shen_integrative_2009}). \\
				$C_2^{tf}$ is derived via nearest shrunken centroid classification.\\
				Comparison of $C_1$ and $C_2^{tf}$ w.r.t.\ their associations with survival. 
			\end{tabular} \\ \midrule
			(Ext) method-based &  \begin{tabular}[t]{@{}l@{}}\textcite{freudenberg_clean_2009},\\ cancer gene expression \end{tabular} & \begin{tabular}[t]{@{}l@{}} Descriptive clustering of cancer-related genes.\\ 
				Two independently collected breast cancer datasets.\\
				Data clustered with CSIMM (\cite{liu_context-specific_2006}).\\ 
				Each gene in the clustering is assigned a score for agreement\\ with functional categories (external measure).\\
				Calculate correlation between the gene scores obtained with $C_1$ and $C_2^{md}$. \end{tabular}\\ \midrule
			% \textcolor{red}{(Vis) result-based} & & \\
			(Vis) method-based & \begin{tabular}[t]{@{}l@{}} \textcite{sweatt_discovery_2019},\\ inflammation in pulmonary\\ arterial hypertension \end{tabular} & \begin{tabular}[t]{@{}l@{}} Inferential clustering of hypertension patients.\\
				Two independently collected proteomic profile datasets.\\ 
				Data clustered with Consensus Clustering (\cite{monti_consensus_2003}).\\
				Clusters of $C_1$ and $C_2^{md}$ were matched manually.\\
				Compared heatmaps and PCA plots for $C_1$ and $C_2^{md}$ \\
				which were generated separately for discovery and validation data,\\
				no common projection space was used.\\
			\end{tabular}\\ \midrule
			(Stab) & \begin{tabular}[t]{@{}l@{}}\textcite{bergstrom_long-term_2001}, \\ spinal pain \end{tabular} & \begin{tabular}[t]{@{}l@{}} Inferential clustering of spinal pain patients. \\
				Two independently collected datasets of spinal pain patients.\\
				Data clustered with $k$-means. \\
				$C_2^{tf}$ is derived via nearest-centroid classification.\\
				The kappa coefficient (a partition similarity index) is used \\ to compare $C_2^{md}$ and $C_2^{tf}.$ 
			\end{tabular} \\
			\bottomrule
		\end{tabular}%}
	\end{table}
\end{landscape}
\restoregeometry
}

All our examples are studies from medicine and health sciences. In other application fields (e.g., market research or social sciences) researchers occasionally use validation data, but most of these studies (and some studies in medicine and health sciences) do not fit into our validation framework. Many studies use the result of the validation on the validation data for method selection (e.g., \textcite{sinclair_performance_2005,jamison_empirically_1988,brennan_womens_2012}). In contrast, we have argued in the introduction and in Section \ref{subsec:stability} that for the purpose of validating a clustering result, method selection should be finished after Step \ref{itm:step1}.

Another validation variant is also frequently found in the literature (e.g, \textcite{phinney_college_2005,kaluza_changing_2000,ailawadi_pursuing_2001,gruber_configurations_2010,homburg_configurations_2008}): method selection is performed on the whole dataset, after which the data is split into two sets. The chosen cluster method is applied to the first set, and then validation on the second set (the validation data) is assessed. Successful ``validation'' may indicate a certain robustness or stability of the result, but again, this approach does not fit into our framework: method selection should be constrained to the first part of the split dataset, and not be performed on the whole data. 

Further, other studies (e.g., \textcite{alexe_data_2006}) perform a procedure that appears similar to method-based validation: They split a dataset into two halves, use the first half as the discovery set, but obtain $C_2^{md}$ by clustering discovery and validation data \textit{together} (instead of only clustering the validation data), which will likely yield a biased validation result.

%\section{Data Example}
%\label{sec:dataexample}
%\textcolor{red}{I have not yet decided which concrete datasets to use here. Possible outline for this section:}
%\begin{itemize}
%	\item perhaps two data examples - one for inferential clustering, one for descriptive clustering (could also be based on the same dataset, see the gene expression example -- clustering patients vs. genes)
%	\item split dataset into half, probably also try to obtain a separate dataset for validation? (more difficult)
%	\item demonstrate that result-based and method-based validation are not ``equivalent'': $C_2^{md}$ could be very different from $C_2^{tf}$ (as mentioned above)
%	\item If possible, I want to give a short example for overoptimism resulting from method selection, because this is one of the main arguments for using a validation set \textit{after} model selection 
%\end{itemize}

\section{Discussion}
\label{sec:discussion}
We have presented a systematic framework for validating clusterings on a validation dataset that encompasses procedures known from the literature as special cases. While we hope that this framework will provide some guidance to researchers, the procedure cannot be performed in an ``automated'' manner. Rather, it requires substantial input from the researchers who must decide which validation criteria are important for them depending on the substantive context. Furthermore, specific indices and plots need to be chosen, as well as whether the amount of agreement between results on the discovery and validation datasets is assessed as sufficient. We have given hints about when some aspects may be of interest, but as every application is different, there are no clear rules. This holds for the clustering process in general: while cluster analysis is often interpreted as being able to find meaningful structure in the data ``on its own'', the choice of cluster concept and method requires thorough consideration by researchers (\cite{hennig_what_2015,akhanli_comparing_2020}). The same is true for our validation framework.

With respect to the choice of validation data, we have explained that a validation dataset could be obtained by splitting the original dataset, or it could be a separately collected dataset. On one hand, if the validation dataset comes from the same distribution as the discovery dataset, a successful validation says nothing about other distributions. Moreover, if the original dataset is split, then this reduces the size of the data and can make it more difficult to find meaningful cluster structure in the data.
%(when using method-based validation, one could select a method on the discovery data, validate it on the validation data, and then give a final solution on both datasets combined). %commented out by Theresa, because this is probably not a solution to the problem. 
On the other hand, if the validation data have been independently collected (potentially coming from a different distribution) and the validation fails, it can be difficult to determine whether this is due to the clustering not being meaningful, or due to systematic differences between discovery and validation data. Conversely, if validation \textit{is} successful, then this is all the more encouraging, because it suggests that the clustering result may be valid in a more general context. 

In conclusion, we hope that the validation of clusterings on validation data---while already prevalent in some fields to a certain extent---will become much more common in the applied literature. As we have discussed in this paper, such validation procedures are of vital importance---after all, no researcher wants to make any claims which cannot be replicated on another dataset.

%\textcolor{red}{Other topics for the discussion section:}
%\begin{itemize}
%	\item After successful validation, one might consider repeating the cluster analysis for the \textit{whole} dataset, and present this result as the ``final'' clustering. In supervised classification one occasionally encounters an analogon: After training the classifier on the training set and validating it on the validation set to obtain an estimate for the prediction error, the researcher then proceeds to train the classifier on the whole dataset and this is the ``final'' classifier. But does this make sense for clustering?
%	\item In the introduction, we explained the difference between our framework and benchmarking studies. There is still a connection between a benchmarking study and validating a clustering result on validation data: One may suspect that if we choose a clustering method that was determined as ``best'' for $\Phi_1$ in a rigorous benchmarking study, then the clustering resulting from this method may be also easier to validate on the validation data. Conversely, an important performance criterion in a benchmarking study could be whether a clustering method yields results that can be validated on validation data. We plan to pursue this line of study in the future. As it stands, there is, in general, a lack of rigorous benchmarking studies for cluster analysis (see \cite{van_mechelen_benchmarking_2018}), leaving the applied researcher who wishes to conduct a cluster analysis to try out several methods in Step~\ref{itm:step1}.	
%\end{itemize}

\section*{Acknowledgments}
We thank Anna Jacob and Alethea Charlton for making valuable language corrections. 

\section*{Funding}
This work has been partially supported by the German Federal Ministry of Education and Research (BMBF) [grant number 01IS18036A to ALB (Munich Center of Machine Learning)] and the German Research Foundation [grant number BO3139/7-1 to ALB]. The authors of this work take full responsibility for its content.

\printbibliography

\end{document}